\documentclass[11pt]{article}

\title{AdS/CFT Correspondence and Unification at About 4 TeV}

\author{Paul H. Frampton\\
University of North Carolina at Chapel Hill}

\date{October 2003}

\begin{document}
\maketitle

\section{Introduction}
 
I will give a brief summary of an approach to string phenomenology
which is inspired by AdS/CFT correspondence and which has been pursued
for the last five years. Finite-N non-SUSY theories as discussed here 
are not obtainable from AdS/CFT although a speculation, currently under study,
is that key UV properties of infinite-N theories which may be so obtained
can survive, at least in some(one?) cases for the finite-N case.
Future work will study the (non-)occurrence of quadratic divergences in
the resultant finite-N gauge theories. 

Independent of the outcome of that study, interesting possibilities emerge
for extending the standard model without low-energy supersymmetry.
These possibilities would become far more compelling if the quadratic 
divergences associated with fundamental scalars can indeed be eliminated.

\section{Quiver Gauge Theory}

The relationship of the Type IIB superstring to conformal gauge theory
in $d=4$ gives rise to an interesting class of gauge
theories.
Choosing the simplest compactification\cite{Maldacena}
on $AdS_5 \times S_5$ gives rise to an ${\cal N} = 4$ SU(N) gauge theory
which is known to be conformal due to
the extended global supersymmetry and non-renormalization theorems. All
of the RGE $\beta-$functions for this ${\cal N} = 4$
case are vanishing in perturbation theory. It is possible to break
the ${\cal N}=4$ to ${\cal N}=2,1,0$ by replacing
$S_5$ by an orbifold $S_5/\Gamma$
where $\Gamma$ is a discrete group with
$\Gamma \subset SU(2), \subset SU(3), \not\subset SU(3)$
respectively.

In building a conformal gauge theory model \cite{Frampton,FS,FV},
the steps are: (1) Choose the discrete group $\Gamma$; (2) Embed
$\Gamma \subset SU(4)$; (3) Choose the $N$ of $SU(N)$; and
(4) Embed the Standard Model $SU(3) \times SU(2) \times U(1)$
in the resultant gauge group $\bigotimes SU(N)^p$ (quiver
node identification). Here we shall look only
at abelian $\Gamma = Z_p$ and define
$\alpha = exp(2 \pi i/p)$. It is expected from the string-field
duality that the resultant field
theory is conformal in the $N\longrightarrow \infty$ limit,
and will have a fixed manifold, or at least a fixed point, for $N$ finite.

Before focusing on ${\cal N}=0$ non-supersymmetric cases, let
us first examine an ${\cal N}=1$ model first
put forward in the work of
Kachru and Silverstein\cite{KS}.
The choice is $\Gamma = Z_3$ and the {\bf 4} of $SU(4)$
is {\bf 4} = $(1, \alpha, \alpha, \alpha^2)$. Choosing N=3
this leads to the three chiral families under $SU(3)^3$
trinification\cite{DGG}
\begin{equation}
(3, \bar{3}, 1) + (1, 3, \bar{3}) + (\bar{3}, 1, 3)
\end{equation}

\section{Gauge Couplings.}

An alternative to conformality, grand unification with supersymmetry,
leads to an impressively accurate gauge coupling unification\cite{ADFFL}.
In particular it predicts an electroweak mixing angle
at the Z-pole, ${\tt sin}^2 \theta = 0.231$. This result
may, however, be fortuitous, but rather than
abandon gauge coupling unification, we can rederive ${\tt sin}^2 \theta = 0.231$
in a different way by embedding the electroweak $SU(2) \times U(1)$ in
$SU(N) \times SU(N) \times SU(N)$ to
find ${\tt sin}^2 \theta = 3/13 \simeq 0.231$\cite{FV,F2}.
This will be a common feature of the models in this paper.

\section{4 TeV Grand Unification}

Conformal invariance in two dimensions has had
great success in comparison to several condensed matter
systems. It is an
interesting question whether conformal symmetry
can have comparable success in a four-dimensional
description of high-energy physics.

Even before the standard model (SM)
$SU(2) \times U(1)$ electroweak theory
was firmly established by experimental
data, proposals were made
\cite{PS,GG} of models which would subsume it into
a grand unified theory (GUT) including also the dynamics\cite{GQW} of
QCD. Although the prediction of
SU(5) in its minimal form for the proton lifetime
has long ago been excluded, {\it ad hoc} variants thereof
\cite{FG} remain viable.
Low-energy supersymmetry improves the accuracy of
unification of the three 321 couplings\cite{ADF,ADFFL}
and such theories encompass a ``desert'' between the weak
scale $\sim 250$ GeV and the much-higher GUT scale
$\sim 2 \times 10^{16}$ GeV, although minimal supersymmetric
$SU(5)$ is by now ruled out\cite{Murayama}.

Recent developments in string theory are suggestive
of a different strategy for unification of electroweak
theory with QCD. Both the desert and low-energy
supersymmetry are abandoned. Instead, the
standard $SU(3)_C \times SU(2)_L \times U(1)_Y$
gauge group is embedded in a semi-simple gauge
group such as $SU(3)^N$ as suggested by gauge
theories arising from compactification of the IIB superstring
on an orbifold $AdS_5 \times S^5/\Gamma$ where
$\Gamma$ is the abelian finite group $Z_N$\cite{Frampton}.
In such nonsupersymmetric quiver gauge theories
the unification of couplings happens not by
logarithmic evolution\cite{GQW} over an
enormous desert covering, say, a dozen orders
of magnitude in energy scale. Instead the
unification occurs abruptly at $\mu = M$ through the
diagonal embeddings of 321 in $SU(3)^N$\cite{F2}.
The key prediction of such unification shifts from
proton decay to additional particle content,
in the present model
at $\simeq 4$ TeV.

Let me consider first the electroweak group
which in the standard model is still un-unified
as $SU(2) \times U(1)$. In the 331-model\cite{PP,PF}
where this is extended to $SU(3) \times U(1)$
there appears a Landau pole at $M \simeq 4$ TeV
because that is the scale at which ${\rm sin}^2
\theta (\mu)$ slides to the value
${\rm sin}^2 (M) = 1/4$.
It is also the scale at which the custodial gauged
$SU(3)$ is broken in the framework
of \cite{DK}.

There remains the question of embedding such unification
in an $SU(3)^N$ of the type described in \cite{Frampton,F2}.
Since the required embedding of $SU(2)_L \times U(1)_Y$
into an $SU(3)$ necessitates $3\alpha_Y=\alpha_H$
the ratios of couplings at $\simeq 4$ TeV
is: $\alpha_{3C} : \alpha_{3W} :  \alpha_{3H} :: 5 : 2 : 2$
and it is natural
to examine $N=12$ with diagonal embeddings of
Color (C), Weak (W) and Hypercharge (H)
in $SU(3)^2, SU(3)^5, SU(3)^5$ respectively.

To accomplish this I specify the embedding
of $\Gamma = Z_{12}$ in the global $SU(4)$ R-parity
of the ${\cal N} = 4$ supersymmetry of the underlying theory.
Defining $\alpha = {\rm exp} ( 2\pi i / 12)$ this specification
can be made by ${\bf 4} \equiv (\alpha^{A_1}, \alpha^{A_2},
\alpha^{A_3}, \alpha^{A_4})$ with $\Sigma A_{\mu} = 0 ({\rm mod} 12)$
and all $A_{\mu} \not= 0$ so that all four supersymmetries are
broken from ${\cal N} = 4$ to ${\cal N} = 0$.

Having specified $A_{\mu}$ I calculate the content
of complex scalars by investigating in $SU(4)$
the ${\bf 6} \equiv (\alpha^{a_1}, \alpha^{a_2}, \alpha^{a_3},
\alpha^{-a_3}, \alpha^{-a_2},\alpha^{-a_1})$ with
$a_1 = A_1 + A_2, a_2 = A_2 + A_3, a_3 = A_3 + A_1$ where
all quantities are defined (mod 12).

Finally I identify the nodes (as C, W or H)
on the dodecahedral quiver such that the complex scalars
\begin{equation}
\Sigma_{i=1}^{i=3} \Sigma_{\alpha=1}^{\alpha=12}
\left( N_{\alpha}, \bar{N}_{\alpha \pm a_i} \right)
\label{scalars2}
\end{equation}
are adequate to allow the required symmetry breaking to the
$SU(3)^3$ diagonal subgroup, and the chiral fermions
\begin{equation}
\Sigma_{\mu=1}^{\mu=4} \Sigma_{\alpha=1}^{\alpha=12}
\left( N_{\alpha}, \bar{N}_{\alpha + A_{\mu}} \right)
\label{fermions2}
\end{equation}
can accommodate the three generations of quarks and leptons.

It is not trivial to accomplish all of these requirements
so let me demonstrate by an explicit example.

For the embedding I take $A_{\mu} = (1, 2, 3, 6)$ and for
the quiver nodes take the ordering:
\begin{equation}
- C - W - H - C - W^4 - H^4 -
\label{quiver}
\end{equation}
with the two ends of (\ref{quiver}) identified.

The scalars follow from $a_i = (3, 4, 5)$
and the scalars in Eq.(\ref{scalars2})
\begin{equation}
\Sigma_{i=1}^{i=3} \Sigma_{\alpha=1}^{\alpha=12}
\left( 3_{\alpha}, \bar{3}_{\alpha \pm a_i} \right)
\label{modelscalars}
\end{equation}
are sufficient to break to all diagonal subgroups as
\begin{equation}
SU(3)_C \times SU(3)_{W} \times SU(3)_{H}
\label{gaugegroup}
\end{equation}

The fermions follow from $A_{\mu}$ in Eq.(\ref{fermions2}) as
\begin{equation}
\Sigma_{\mu=1}^{\mu=4} \Sigma_{\alpha=1}^{\alpha=12}
\left( 3_{\alpha}, \bar{3}_{\alpha + A_{\mu}} \right)
\label{modelfermions}
\end{equation}
and the particular dodecahedral quiver
in (\ref{quiver}) gives rise  to exactly {\it three}
chiral generations which transform under (\ref{gaugegroup})
as
\begin{equation}
3[ (3, \bar{3}, 1) + (\bar{3}, 1, 3) + (1, 3, \bar{3}) ]
\label{generations}
\end{equation}
I note that anomaly freedom of the underlying superstring
dictates that only the combination
of states in Eq.(\ref{generations})
can survive. Thus, it
is sufficient to examine one of the terms, say
$( 3, \bar{3}, 1)$. By drawing the quiver diagram
indicated by Eq.(\ref{quiver}) with the twelve nodes
on a ``clock-face'' and using
$A_{\mu} = (1, 2, 3, 6)$
I find
five $(3, \bar{3}, 1)$'s and two $(\bar{3}, 3, 1)$'s
implying three chiral families as stated in Eq.(\ref{generations}).

After further symmetry breaking at scale $M$ to
$SU(3)_C \times SU(2)_L \times U(1)_Y$ the
surviving chiral fermions are the quarks and leptons
of the SM. The appearance
of three families depends on both
the identification of modes in (\ref{quiver})
and on the embedding of $\Gamma \subset SU(4)$. The
embedding must simultaneously give adequate
scalars whose VEVs can break the symmetry
spontaneously to (\ref{gaugegroup}).
All of this is achieved successfully by the
choices made.
The three gauge couplings evolve
for $M_Z \leq \mu \leq M$. For $\mu \geq M$ the
(equal) gauge couplings of $SU(3)^{12}$
do not run if, as conjectured in \cite{Frampton,F2}
there is a conformal fixed point at $\mu = M$.

The basis of the conjecture in \cite{Frampton,F2}
is the proposed duality of Maldacena\cite{Maldacena}
which shows that in the $N \rightarrow \infty$
limit ${\cal N} = 4$ supersymmetric
$SU(N)$gauge  theory, as well as orbifolded versions with
${\cal N} = 2,1$ and $0$\cite{bershadsky1,bershadsky2}
become conformally invariant.
It was known long ago
that the
${\cal N} = 4$ theory is
conformally invariant for all finite ${\cal N} \geq 2$.
This led to the conjecture in \cite{Frampton}
that the ${\cal N} = 0$
theories might be conformally
invariant, at least in some case(s),
for finite $N$.
It should be emphasized that this
conjecture cannot be checked
purely
within a perturbative framework\cite{FMink}.
I assume that the local $U(1)$'s
which arise in this scenario
and which would lead to $U(N)$
gauge groups are non-dynamical,
as suggested by Witten\cite{Witten},
leaving $SU(N)$'s.

As for experimental tests of such
a TeV GUT, the situation at energies
below 4 TeV is predicted to be the standard model with
a Higgs boson still to be discovered at a mass
predicted by radiative corrections
\cite{PDG} to be below 267 GeV at 99\% confidence level.

There are many particles predicted
at $\simeq 4$ TeV beyond those of
the minimal standard model.
They include
as spin-0 scalars the states of Eq.(\ref{modelscalars}).
and
as spin-1/2 fermions the states
of Eq.(\ref{modelfermions}),
Also predicted are gauge bosons to fill out the gauge groups
of (\ref{gaugegroup}), and in the same energy region
the gauge bosons to fill out all of
$SU(3)^{12}$. All these extra particles are necessitated by
the conformality constraints of \cite{Frampton,F2} to lie
close to the conformal fixed point.

One important issue is whether this proliferation
of states at $\sim 4$ TeV is
compatible with precision
electroweak data in hand. This has
been studied in the related model of
\cite{DK} in a recent article\cite{Csaki}. Those results
are not easily translated to the present
model but it is possible that such an analysis
including limits on flavor-changing neutral currents
could rule out the entire framework.

\section{Predictivity}

The calculations have been done in the one-loop
approximation to the renormalization group equations
and threshold effects have been ignored.
These corrections are not expected to be large
since the couplings are weak in the entrire energy
range considered. There are possible further corrections
such a non-perturbative effects, and the effects of
large extra dimensions, if any.

In one sense the robustness of this TeV-scale
unification is almost self-evident, in that it follows from the weakness
of the coupling constants in the evolution from $M_Z$ to $M_U$.
That is, in order to define the theory at $M_U$,
one must combine the effects of
threshold corrections ( due to O($\alpha(M_U)$)
mass splittings )
and potential corrections from redefinitions
of the coupling constants and the unification scale.
We can then {\it impose} the coupling constant relations at $M_U$
as renormalization conditions and this is valid
to the extent that higher order corrections do
not destabilize the vacuum state.

We shall approach the comparison with data in two
different but almost equivalent ways. The first
is "bottom-up" where we use as input that the
values of $\alpha_3(\mu)/\alpha_2(\mu)$ and
$\sin^2 \theta (\mu)$ are expected to be $5/2$
and $1/4$ respectively at $\mu = M_U$.

Using the experimental ranges allowed for
$\sin^2 \theta (M_Z) = 0.23113 \pm 0.00015$,
$\alpha_3 (M_Z) = 0.1172 \pm 0.0020$ and
$\alpha_{em}^{-1} (M_Z) = 127.934 \pm 0.027$
\cite{PDG} we have calculated \cite{FRT}
the values of $\sin^2 \theta (M_U)$
and $\alpha_3 (M_U) / \alpha_2(M_U)$
for a range of $M_U$ between 1.5 TeV
and 8 TeV.
Allowing a maximum discrepancy of $\pm 1\%$ in
$\sin^2 \theta (M_U)$ and
$\pm 4\%$ in $\alpha_3 (M_U) / \alpha_2 (M_U)$
as reasonable estimates of corrections, we deduce that
the unification scale $M_U$ can lie anywhere
between 2.5 TeV and 5 TeV. Thus the theory is
robust in the sense that there is no singular
limit involved in choosing a particular value
of $M_U$.

\bigskip

\noindent Another test of predictivity of the same
model is to fix the unification values at $M_U$ of
$\sin^2 \theta(M_U) = 1/4$ and $\alpha_3 (M_U) /
\alpha_2 (M_U) = 5/2$. We then compute the
resultant predictions at the scale $\mu = M_Z$.

The results are shown for $\sin^2 \theta (M_Z)$
in \cite{FRT} with the allowed range\cite{PDG}
$\alpha_3 (M_Z) = 0.1172 \pm 0.0020$. The precise
data on $\sin^2 (M_Z)$ are indicated in \cite{FRT} and
the conclusion is that the model makes correct
predictions for $\sin^2 \theta (M_Z)$.
Similarly, in \cite{FRT}, there is a plot of the
prediction for $\alpha_3 (M_Z)$ versus
$M_U$ with $\sin^2 \theta(M_Z)$ held
with the allowed empirical range.
The two quantities plotted in \cite{FRT}
are consistent for similar ranges of $M_U$.
Both $\sin^2 \theta(M_Z)$ and $\alpha_3(M_Z)$ are within the empirical limits
if $M_U = 3.8 \pm 0.4$ TeV.

\bigskip

\noindent The model has many additional gauge bosons
at the unification scale, including neutral $Z^{'}$'s,
which could mediate flavor-changing processes
on which there are strong empirical upper limits.

A detailed analysis wll require specific identification
of the light families and quark flavors with
the chiral fermions appearing in the quiver diagram
for the model. We can make only the general
observation that the lower bound on a $Z^{'}$
which couples like the standard $Z$ boson
is quoted as $M(Z^{'}) < 1.5$ TeV \cite{PDG}
which is safely below the $M_U$ values considered here
and which we identify with the mass of the new gauge bosons.

This is encouraging to believe that flavor-changing
processes are under control in the model but
this issue will require more careful analysis when
a specific identification of the quark states
is attempted.

\bigskip

\noindent Since there are many new states predicted
at the unification scale $\sim 4$ TeV, there is a danger
of being ruled out by precision low energy data.
This issue is conveniently studied in terms
of the parameters $S$ and $T$ introduced in \cite{Peskin}
and designed to measure departure from the predictions
of the standard model.

Concerning $T$, if the new $SU(2)$ doublets are
mass-degenerate and hence do not violate a custodial
$SU(2)$ symmetry they contribute nothing to $T$.
This therefore provides a constraint on the spectrum
of new states.

\section{Discussion}

\bigskip

\noindent The plots we have presented clarify the accuracy
of the predictions of this TeV unification scheme for
the precision values accurately measured at the Z-pole.
The predictivity is as accurate for $\sin^2 \theta$ as
it is for supersymmetric GUT models\cite{ADFFL,ADF,DRW,DG}.
There is, in addition, an accurate prediction for $\alpha_3$
which is used merely as input in SusyGUT models.

At the same time, the accurate predictions are seen to be robust
under varying the unification scale around $\sim 4 TeV$
from about 2.5 TeV to 5 TeV.

One interesting question is concerning the accommodation
of neutrino masses in view of the popularity of the mechanisms
which require a higher mass scale than occurs in the present
type of model. For example, one would like to
know whether any of the recent studies in \cite{FGMY}
can be useful within this framework.

In conclusion, since this model ameliorates the GUT hierarchy
problem and naturally accommodates three families, it
provides a viable alternative to the widely-studied
GUT models which unify by logarithmic evolution
of couplings up to much higher GUT scales.

\section*{Acknowledgement}

\bigskip
\bigskip
I thank  Professor N. Mankoc-Borstnik of the University
of Ljubjana
for organizing this stimulating workshop in Portoroz,
Slovenia.
This work was supported in part
by the Department of Energy
under Grant Number
DE-FG02-97ER-410236.


\bigskip
\bigskip

\begin{thebibliography}{99}
\bibitem{Maldacena}
J. Maldacena, Adv. Theor. Math. Phys. {\bf 2,} 231 (1998).
{\tt hep-th/9711200}.\\
S.S. Gubser, I.R. Klebanov and A.M. Polyakov, Phys. Lett.
{\bf B428,} 105 (1998). {\tt hep-th/9802109}.\\
E. Witten, Adv. Theor. Math. Phys. {\bf 2,} 253 (1998).
{\tt hep-th/9802150}.
\bibitem{Frampton}
P.H. Frampton, Phys. Rev. {\bf D60,} 041901 (1999). {\tt hep-th/9812117}.
\bibitem{FS}
P.H. Frampton and W.F. Shively, Phys. Lett. {\bf B454,} 49 (1999).
{\tt hep-th/9902168}.
\bibitem{FV}
P.H. Frampton and C. Vafa. {\tt hep-th/9903226.}
\bibitem{KS}
S. Kachru and E. Silverstein, Phys. Rev. Lett. {\bf 80,} 4855 (1998).
{\tt hep-th/9802183}.
\bibitem{DGG}
A. De R\'{u}jula, H. Georgi and S.L. Glashow.
{\it Fifth Workshop on Grand Unification}.
Editors: P.H. Frampton, H. Fried and K.Kang.
World Scientific (1984) page 88.
\bibitem{ADFFL}
U. Amaldi, W. De Boer, P.H. Frampton, H. F\"{u}rstenau and J.T. Liu.
Phys. Lett. {\bf B281,} 374 (1992).
\bibitem{F2}
P.H. Frampton, Phys. Rev. {\bf D60,} 085004 (1999). {\tt hep-th/9905042}.
\bibitem{PS}
J.C. Pati and A. Salam,
Phys. Rev. {\bf D8,} 1240 (1973); {\it ibid} {\bf D10,} 275 (1974).
\bibitem{GG}
H. Georgi and S.L. Glashow, Phys. Rev. Lett. {\bf 32,} 438 (1974).
\bibitem{GQW}
H. Georgi, H.R. Quinn and S. Weinberg, Phys. Rev. Lett.
{\bf 33,} 451 (1974).
\bibitem{FG}
P.H. Frampton and S.L. Glashow, Phys. Lett. {\bf B131,} 340 (1983).
\bibitem{ADF}
U. Amaldi, W. de Boer and H. F\"{u}rstenau, Phys. Lett. {\bf B260,} 447 (1991).
\bibitem{Murayama}
H. Murayama and A. Pierce, Phys. Rev. {\bf D65,}
055009 (2002).
\bibitem{PP}
F. Pisano and V. Pleitez, Phys. Rev. {\bf D46,} 410 (1992).
\bibitem{PF}
P.H. Frampton,
Phys. Rev. Lett. {\bf 69,} 2889 (1992).
\bibitem{DK}
S. Dimopoulos and D. E. Kaplan, Phys. Lett. {\bf B531,} 127 (2002).
\bibitem{bershadsky1}
M. Bershadsky, Z. Kakushadze and C. Vafa,
Nucl. Phys. {\bf B523,} 59 (1998).
\bibitem{bershadsky2}
M. Bershadsky and A. Johansen, Nucl. Phys. {\bf B536,} 141 (1998).
\bibitem{FMink}
P.H. Frampton and P. Minkowski, {\tt hep-th/0208024}.
\bibitem{Witten}
E. Witten, JHEP {\bf 9812:012} (1998).
\bibitem{PDG}
Particle Data Group. {\it Review of Particle Physics.}
Phys. Rev. {\bf D66,} 010001 (2002).
\bibitem{Csaki}
C. Csaki, J. Erlich, G.D. Kribs and J. Terning,
Phys. Rev. {\bf D66,} 075008 (2002).
\bibitem{FRT}
P.H. Frampton, R.M. Rohm and T. Takahashi. Phys. Lett. {\bf B570,} 67 (2003).
{\tt hep-ph/0302074}.
\bibitem{Peskin}
M. Peskin and T. Takeuchi, Phys. Rev. Lett. {\bf 65,} 964 (1990);
Phys. Rev. {\bf D46,} 381 (1992).
\bibitem{DRW}
S. Dimopoulos, S. Raby and F. Wilczek, Phys. Rev. {\bf D24,} 1681 (1981);
Phys. Lett. {\bf B112,} 133 (1982).
\bibitem{DG}
S. Dimopoulos and H. Georgi, Nucl. Phys. {\bf B193,} 150 (1981).
\bibitem{FGMY}
P.H. Frampton and S.L. Glashow, Phys. Lett. {\bf B461,} 95 (1999).
P.H. Frampton, S.L. Glashow and D. Marfatia, Phys. Lett. {\bf B536,} 79 (2002).
P.H. Frampton, S.L. Glashow and T. Yanagida,  Phys. Lett. {\bf B548,} 119 (2002).

\end{thebibliography}
\end{document}